\documentstyle[12pt]{article}

\input epsf

\ifx\epsffile\undefined
\newlength{\epsfysize}
\def\epsffile#1#2#3#4]#5{}
\else
\fi

\catcode`\@=11
\long\def\@makefntext#1{
\protect\noindent \hbox to 3.2pt {\hskip-.9pt
$^{{\ninerm\@thefnmark}}$\hfil}#1\hfill}		

 \def\@makefnmark{\hbox to 0pt{$^{\@thefnmark}$\hss}}  

\def\ps@myheadings{\let\@mkboth\@gobbletwo
\def\@oddhead{\hbox{}
\rightmark\hfil\ninerm\thepage}
\def\@oddfoot{}\def\@evenhead{\ninerm\thepage\hfil
\leftmark\hbox{}}\def\@evenfoot{}
\def\sectionmark##1{}\def\subsectionmark##1{}}

\newcounter{sectionc}\newcounter{subsectionc}\newcounter{subsubsectionc}
\renewcommand{\section}[1] {\vspace{0.6cm}\addtocounter{sectionc}{1}
\setcounter{subsectionc}{0}\setcounter{subsubsectionc}{0}\noindent
	{\bf\thesectionc. #1}\par\vspace{0.4cm}}
\renewcommand{\subsection}[1] {\vspace{0.6cm}\addtocounter{subsectionc}{1}
	\setcounter{subsubsectionc}{0}\noindent
	{\it\thesectionc.\thesubsectionc. #1}\par\vspace{0.4cm}}
\renewcommand{\subsubsection}[1]
{\vspace{0.6cm}\addtocounter{subsubsectionc}{1}
	\noindent {\rm\thesectionc.\thesubsectionc.\thesubsubsectionc.
	#1}\par\vspace{0.4cm}}

\newcounter{appendixc}
\newcounter{subappendixc}[appendixc]
\newcounter{subsubappendixc}[subappendixc]

\renewcommand{\appendix}[1] {\vspace{0.6cm}
        \refstepcounter{appendixc}
        \setcounter{figure}{0}
        \setcounter{table}{0}
        \setcounter{equation}{0}
        \renewcommand{\thefigure}{\Alph{appendixc}.\arabic{figure}}
        \renewcommand{\thetable}{\Alph{appendixc}.\arabic{table}}
        \renewcommand{\theappendixc}{\Alph{appendixc}}
        \renewcommand{\theequation}{\Alph{appendixc}.\arabic{equation}}
        \noindent{\bf Appendix \theappendixc #1}\par\vspace{0.4cm}}

\def\abstracts#1{{
	\centering{\begin{minipage}{30pc}\tenrm\baselineskip=12pt\noindent
	\parindent=0pt #1
	\end{minipage}}\par}}

\renewenvironment{thebibliography}[1]
	{\begin{list}{\arabic{enumi}.}
	{\usecounter{enumi}\setlength{\parsep}{0pt}
\setlength{\leftmargin 1.25cm}{\rightmargin 0pt}
	 \setlength{\itemsep}{0pt} \settowidth
	{\labelwidth}{#1.}\sloppy}}{\end{list}}

\newcounter{itemlistc}
\newcounter{romanlistc}
\newcounter{alphlistc}
\newcounter{arabiclistc}

\newcommand{\fcaption}[1]{
        \refstepcounter{figure}
        \setbox\@tempboxa = \hbox{\tenrm Fig.~\thefigure. #1}
        \ifdim \wd\@tempboxa > 5.5in
           {\begin{center}
        \parbox{5.5in}{\footnotesize\baselineskip=12pt Fig.~\thefigure. #1}
            \end{center}}
        \else
             {\begin{center}
             {\tenrm Fig.~\thefigure. #1}
              \end{center}}
        \fi}

\newcommand{\tcaption}[1]{
        \refstepcounter{table}
        \setbox\@tempboxa = \hbox{\tenrm Table~\thetable. #1}
        \ifdim \wd\@tempboxa > 6in
           {\begin{center}
        \parbox{6in}{\tenrm\baselineskip=12pt Table~\thetable. #1}
            \end{center}}
        \else
             {\begin{center}
             {\tenrm Table~\thetable. #1}
              \end{center}}
        \fi}

\def\@citex[#1]#2{\if@filesw\immediate\write\@auxout
	{\string\citation{#2}}\fi
\def\@citea{}\@cite{\@for\@citeb:=#2\do
	{\@citea\def\@citea{,}\@ifundefined
	{b@\@citeb}{{\bf ?}\@warning
	{Citation `\@citeb' on page \thepage \space undefined}}
	{\csname b@\@citeb\endcsname}}}{#1}}

\newif\if@cghi

\def\fnt#1#2{\footnotetext{\kern-.3em
	{$^{\mbox{\sevenrm #1}}$}{#2}}}

 1
 1
 1

\font\tenbf=cmbx10
\font\tenrm=cmr10
\font\tenit=cmti10

\font\ninerm=cmr9

\def\M{{\cal M}}

\begin{document}

\centerline{\tenbf PRECISE PREDICTIONS FOR MASSES AND COUPLINGS}
\baselineskip=16pt
\centerline{\tenbf IN THE MINIMAL SUPERSYMMETRIC STANDARD MODEL}
\vspace{0.8cm}
\centerline{\tenrm J.~BAGGER, K.~MATCHEV AND D.~PIERCE}
\baselineskip=13pt
\centerline{\tenit Department of Physics and Astronomy}
\baselineskip=12pt
\centerline{\tenit The Johns Hopkins University}
\centerline{\tenit Baltimore, MD  \ 21218,\ \ USA}
\vspace{0.9cm}
\abstracts{
We present selected results of our program to determine the
masses, gauge couplings, and Yukawa couplings of the minimal
supersymmetric model in a full one-loop calculation.  We focus
on the precise prediction of the strong coupling $\alpha_s(M_Z)$
in the context of supersymmetric unification.  We
discuss the importance of including the finite corrections and
demonstrate that the leading-logarithmic approximation can significantly
underestimate $\alpha_s(M_Z)$ when some superpartner masses are light.
We show that if GUT thresholds are ignored, and the superpartner
masses are less than about 500 GeV, the prediction for $\alpha_s(M_Z)$
is quite large.  We impose constraints from nucleon decay experiments
and find that minimal SU(5) GUT threshold corrections increase
$\alpha_s(M_Z)$ over most of the parameter space.  We also consider
the missing-doublet SU(5) model and find that it predicts preferred
values for the strong coupling, even for a very light superpartner
spectrum.  We briefly discuss predictions for the bottom-quark mass
in the small $\tan\beta$ region.}
\vfil

\def\o{\over}
\def\dr{\mbox{\footnotesize$\overline{\it DR}$}~}
\def\ms{\mbox{\footnotesize$\overline{\it MS}$}~}
\def\roughly#1{\raise.3ex\hbox{$#1$\kern-.75em\lower1ex\hbox{$\sim$}}}

\rm\baselineskip=14pt
\section{Introduction}
\vspace{-.25cm}

The exact one-loop corrections to the masses, gauge couplings and
Yukawa couplings of the minimal supersymmetric model are described
in Ref.~\cite{bmp}.  These corrections are essential ingredients
for accurate tests of grand unification.  They allow one to extract
the underlying \dr parameters from a given set of measured observables.
The \dr parameters can then be run up to a high scale to explore
the consequences of different unification hypotheses.

Alternatively, the radiative corrections can be used to translate
various limits into excluded regions of the \dr parameter  space.  This
is illustrated in Fig.~1, where we show the excluded region of the $M_0,
\ M_{1/2}$ parameter space at the tree and one-loop levels, from current
experimental constraints.\footnote{$M_0$ is the universal scalar mass,
$M_{1/2}$ is the universal gaugino mass, and $A_0$ is the universal
$A$-term.}
Here, ``tree-level" means that the
superpartner masses are determined from the \dr parameters evaluated
at the scale $M_Z$.  For the Higgs mass, the tree-level curve corresponds
to the one-loop mass neglecting the gauge/Higgs/gaugino/Higgsino
contributions.

\begin{figure}[t]
\epsfysize=3in
\epsffile[30 495 600 755]{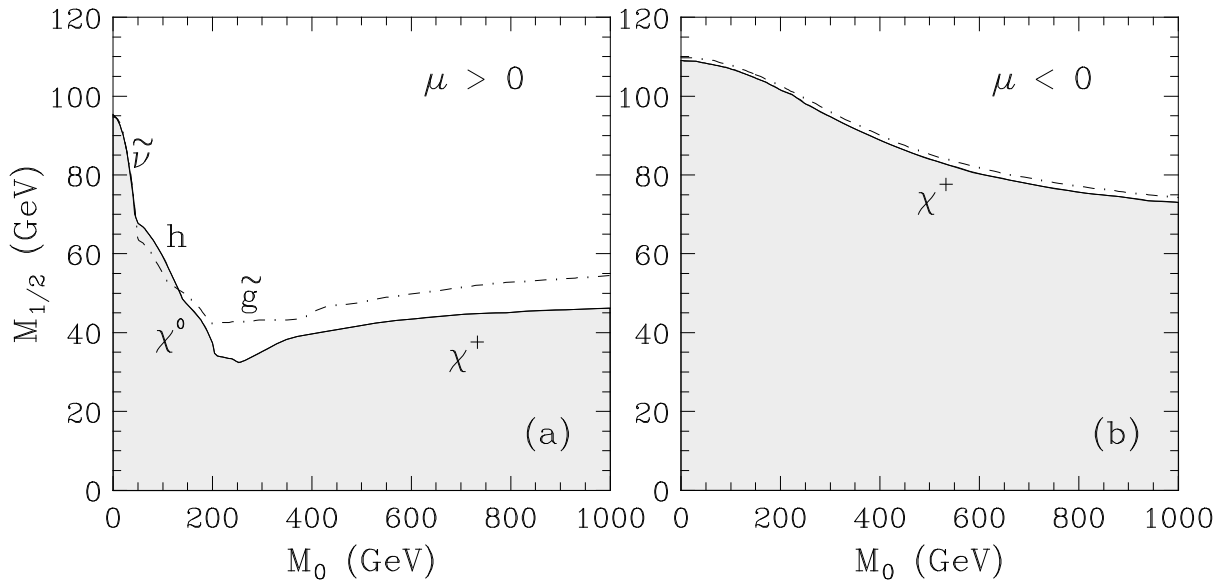}
\begin{center}
\parbox{5.5in}{
\caption[]{\small Excluded region (shaded)
of the $M_0,\ M_{1/2}$ plane,
for $\tan\beta=2,\ m_t=170$ GeV, and $A_0=0$.
All masses are evaluated at one-loop.
The symbols indicate which experimental constraint is relevant:
$\chi^+\Rightarrow m_{\chi^+} > 47$ GeV; $\tilde g\Rightarrow
m_{\tilde g}>125$ GeV; $\tilde\nu\Rightarrow m_{\tilde\nu}>42$ GeV;
$h\Rightarrow m_h>58$ GeV; $\chi^0\Rightarrow$ direct searches for
neutralinos at LEP. The dashed line shows
the boundary of the excluded region at tree-level.}}
\end{center}
\end{figure}

\begin{figure}[t]
\epsfysize=3in
\epsffile[30 468 600 728]{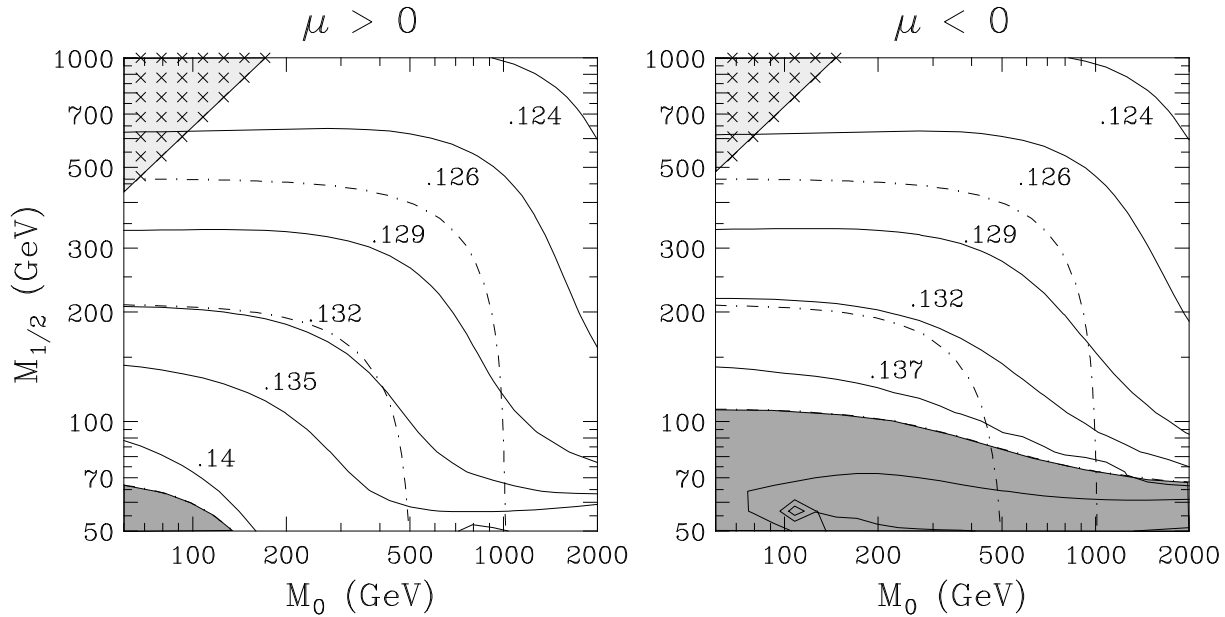}
\begin{center}
\parbox{5.5in}{
\caption[]{\small Contours of $\alpha_s(M_Z)$ in the $M_0$, $M_{1/2}$
plane, with $\tan\beta=2$, $m_t=170$ GeV, and $A_0=0$. The contours
of squark mass 500 and 1000 GeV are shown by dot-dashed lines. The
$\times$'s in the upper left hand corners indicate regions in which
the tau sneutrino is the LSP. The shaded regions are excluded by
particle searches.}}
\end{center}
\end{figure}

\begin{figure}[t]
\epsfysize=3in
\epsffile[50 428 600 698]{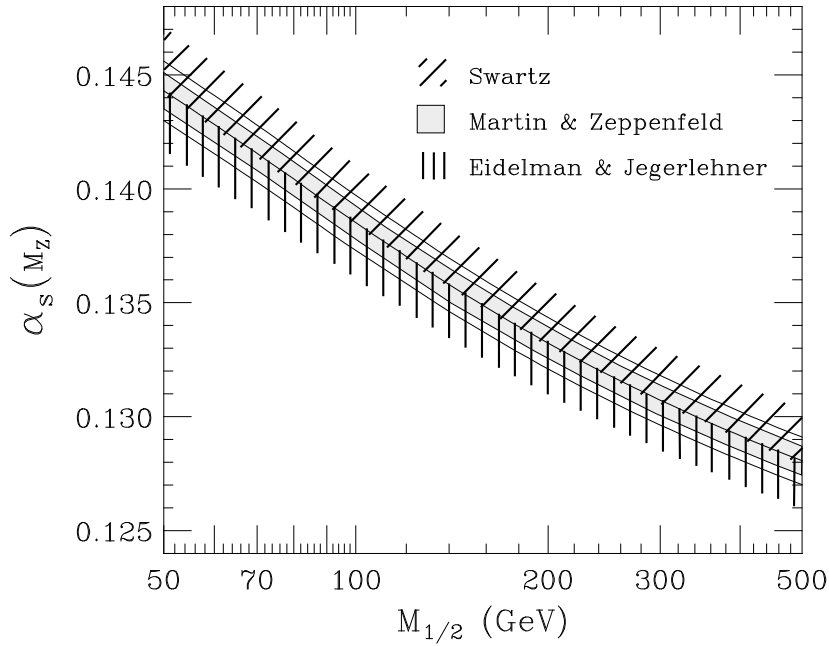}
\begin{center}
\parbox{5.5in}{
\caption[]{\small Predictions for $\alpha_s(M_Z)$ using three different
values for the QED coupling $\alpha(M_Z)$. In each case we show the
$\pm1$--$\sigma$ bands, for $\tan\beta=2,\ m_t=170$ GeV, $A_0=0$,
and $\mu>0$.}}
\end{center}
\end{figure}

\begin{figure}[t]
\epsfysize=3in
\epsffile[35 475 600 735]{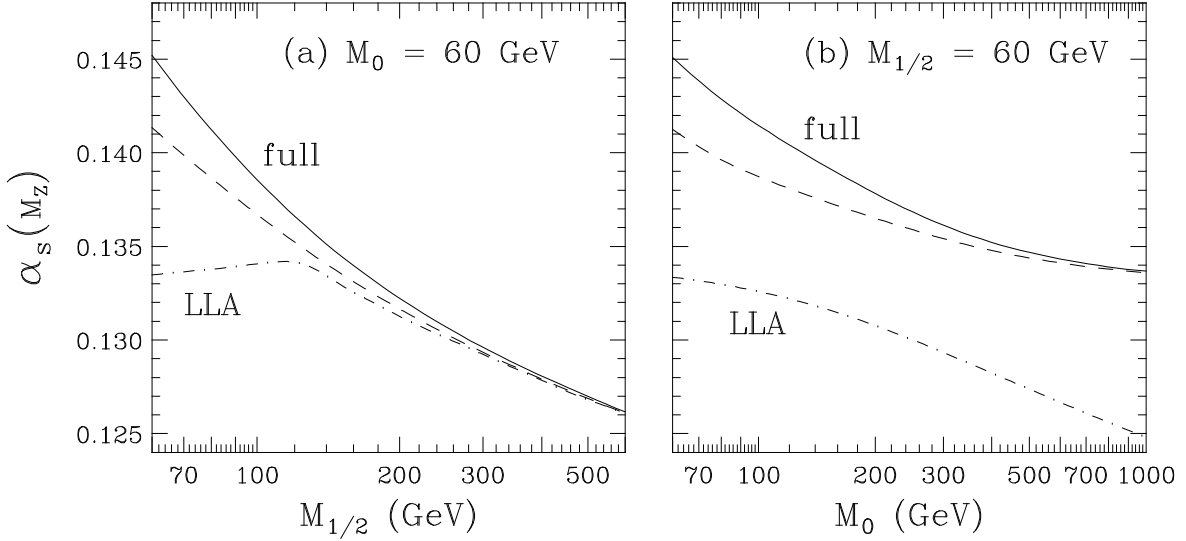}
\begin{center}
\parbox{5.5in}{
\caption[]{\small Comparison of $\alpha_s(M_Z)$ in the leading logarithm
approximation (LLA) versus the full one-loop calculation, for $\tan\beta=2,
\ m_t=170$ GeV, $A_0=0$, and $\mu>0$. The dashed line shows the result if
the non-universal corrections are neglected.}}
\end{center}
\end{figure}

\begin{figure}[t]
\epsfysize=3in
\epsffile[38 475 600 735]{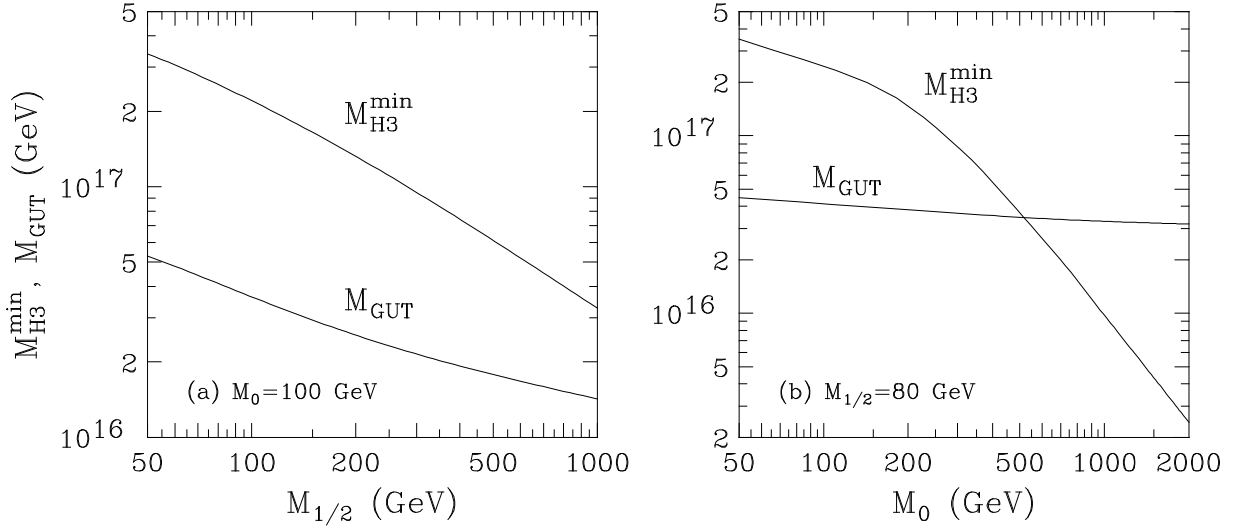}
\begin{center}
\parbox{5.5in}{
\caption[]{\small The smallest triplet Higgs mass allowed by nucleon
decay constraints in minimal SU(5), with $\tan\beta=2,\ m_t=$170 GeV,
$A_0=0$, and $\mu>0$. The GUT scale is also shown.}}
\end{center}
\end{figure}

\begin{figure}[t]
\epsfysize=3in
\epsffile[30 468 600 728]{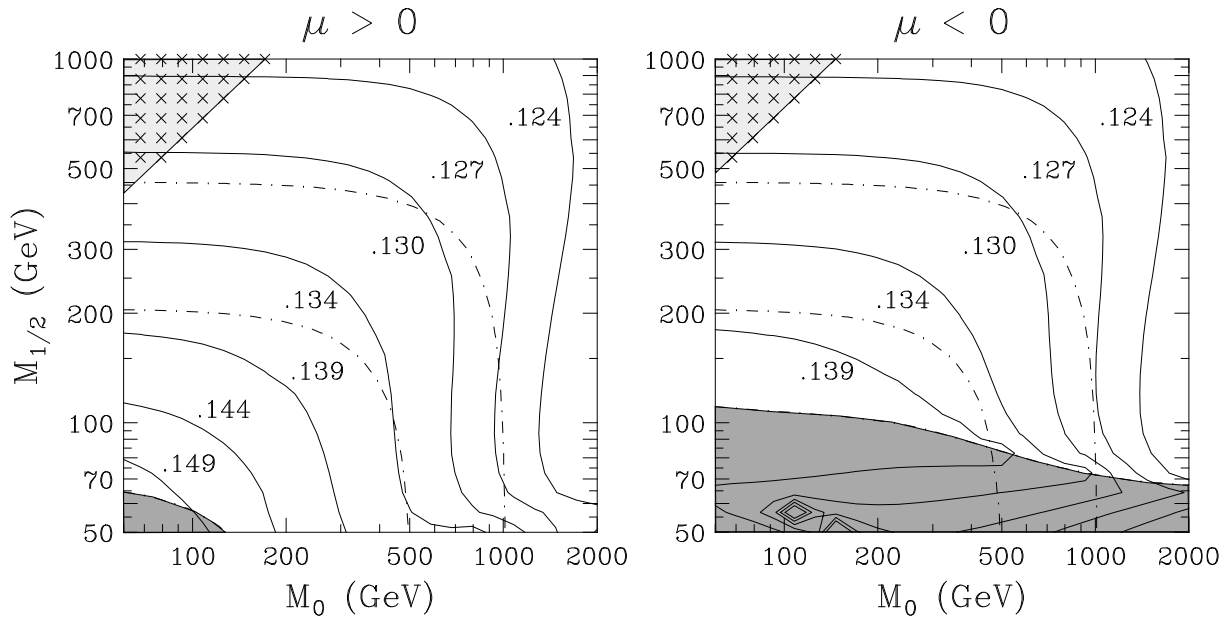}
\begin{center}
\parbox{5.5in}{
\caption[]{\small Contours of the smallest possible $\alpha_s(M_Z)$ consistent
with nucleon decay in minimal SU(5), with
$\tan\beta=2,\ m_t=170$ GeV, and $A_0=0$. The dot-dashed lines indicate
contours of 500 and 1000 GeV squark masses. The $\times$'s indicate the region
where the LSP is charged, and the lower shaded regions are excluded by
particle searches.}}
\end{center}
\end{figure}

\begin{figure}[t]
\epsfysize=3in
\epsffile[30 468 600 728]{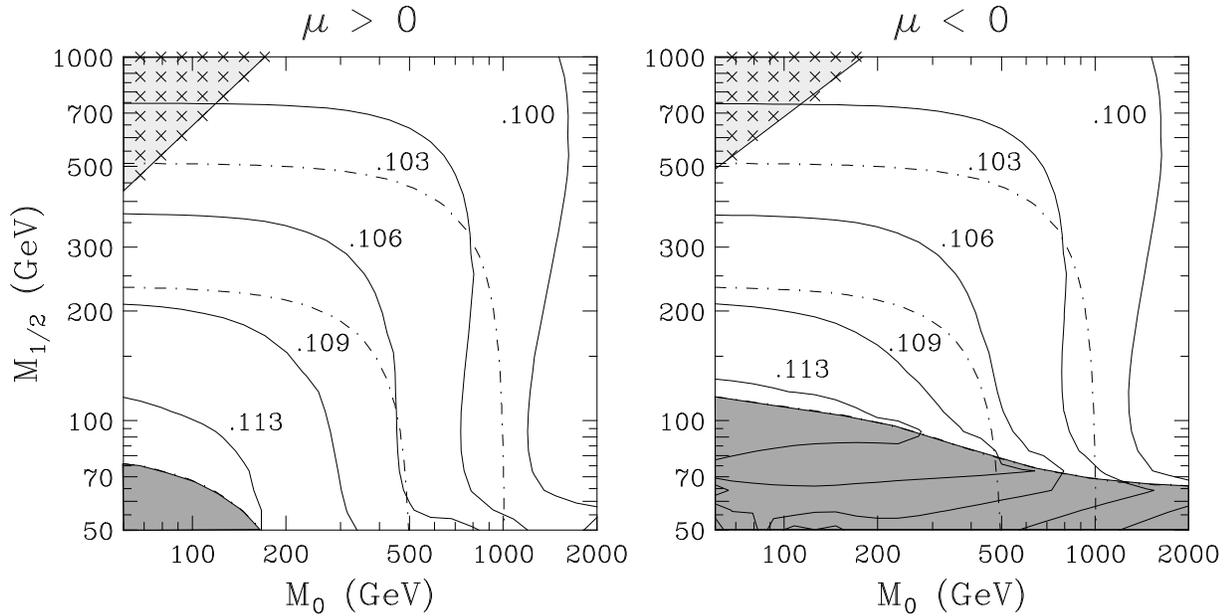}
\begin{center}
\parbox{5.5in}{
\caption[]{\small Contours of the minimum possible $\alpha_s(M_Z)$ consistent
with nucleon decay in the missing-doublet model, with
$\tan\beta=2,\ m_t$=170 GeV, and $A_0=0$. The shading and dashed lines
are as in Fig.~6.}}
\end{center}
\end{figure}

\begin{figure}[t]
\epsfysize=3in
\epsffile[30 465 600 755]{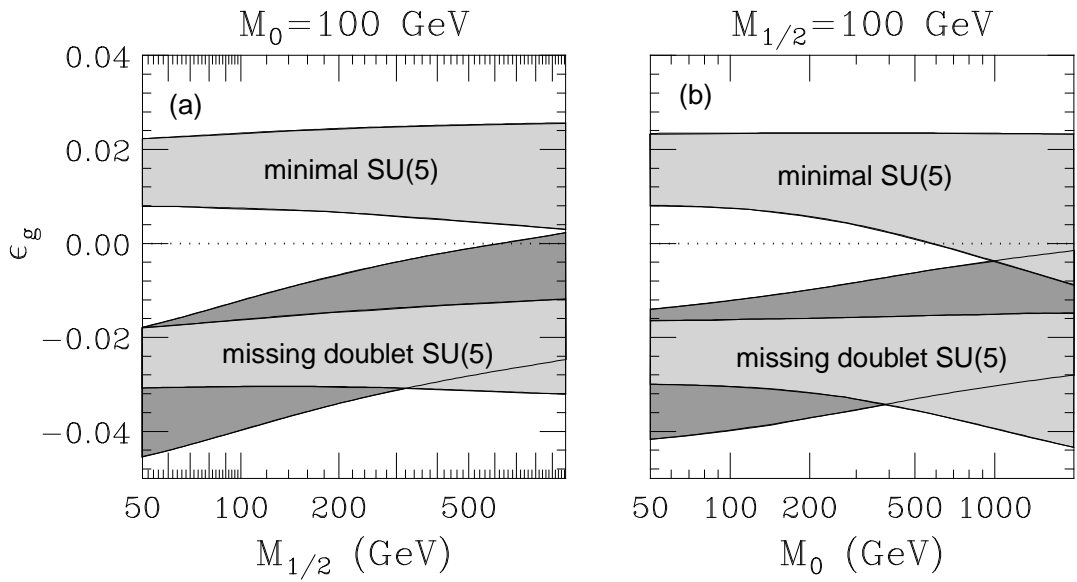}
\begin{center}
\parbox{5.5in}{
\caption[]{\small The light shaded regions indicate the allowed
values of the gauge coupling threshold correction $\varepsilon_g$
in the minimal and missing-doublet SU(5) models.
The dark shaded region indicates the range of $\varepsilon_g$
necessary to obtain $\alpha_s(M_Z) = 0.117\pm0.01$. For $\tan\beta=2,
\ m_t=170$ GeV, $A_0=0$, and $\mu>0$. (From Ref.~\cite{bmp plb}.)}}
\end{center}
\end{figure}

\begin{figure}[t]
\epsfysize=3.5in
\epsffile[25 398 600 718]{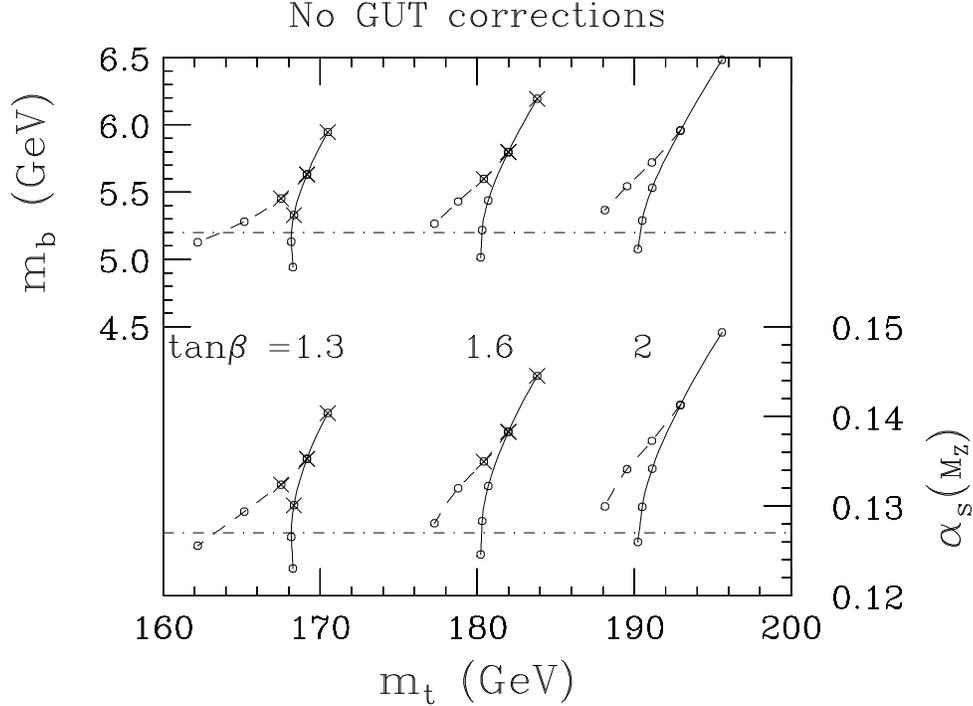}
\begin{center}
\parbox{5.5in}{
\caption[]{\small The bottom-quark mass and $\alpha_s(M_Z)$
vs. $m_t$ for the case of no GUT-scale thresholds,
for various values of $\tan\beta$, with $A_0=0$, $\mu>0$, and
$\hat\lambda_t(M_{\rm GUT})=3$. The right (solid) leg in
each pair of lines corresponds
to $M_{1/2}$ varying from 60 to 1000 GeV, with $M_0$ fixed at 60 GeV.
The left (dashed) leg corresponds to $M_0$ varying from
60 to 1000 GeV, with $M_{1/2}=100$ GeV. On the solid lines the circles
mark, from top to bottom, $M_{1/2}=60$, 100, 200, 400, and 1000 GeV,
and on the dashed lines the circles mark $M_0=60$, 200, 400, and
1000 GeV. Note that the lowest point on each left leg and the
second-to-lowest point on each right leg corresponds to
$m_{\tilde q}\simeq 1$ TeV. The horizontal dashed lines indicate
$m_b=5.2$ GeV and $\alpha_s(M_Z)=0.127$. The $\times$'s mark points
with one-loop Higgs mass $m_h<60$ GeV. (From Ref.~\cite{bmp plb}.)}}
\end{center}
\end{figure}

\begin{figure}[t]
\epsfysize=3.5in
\epsffile[30 405 600 725]{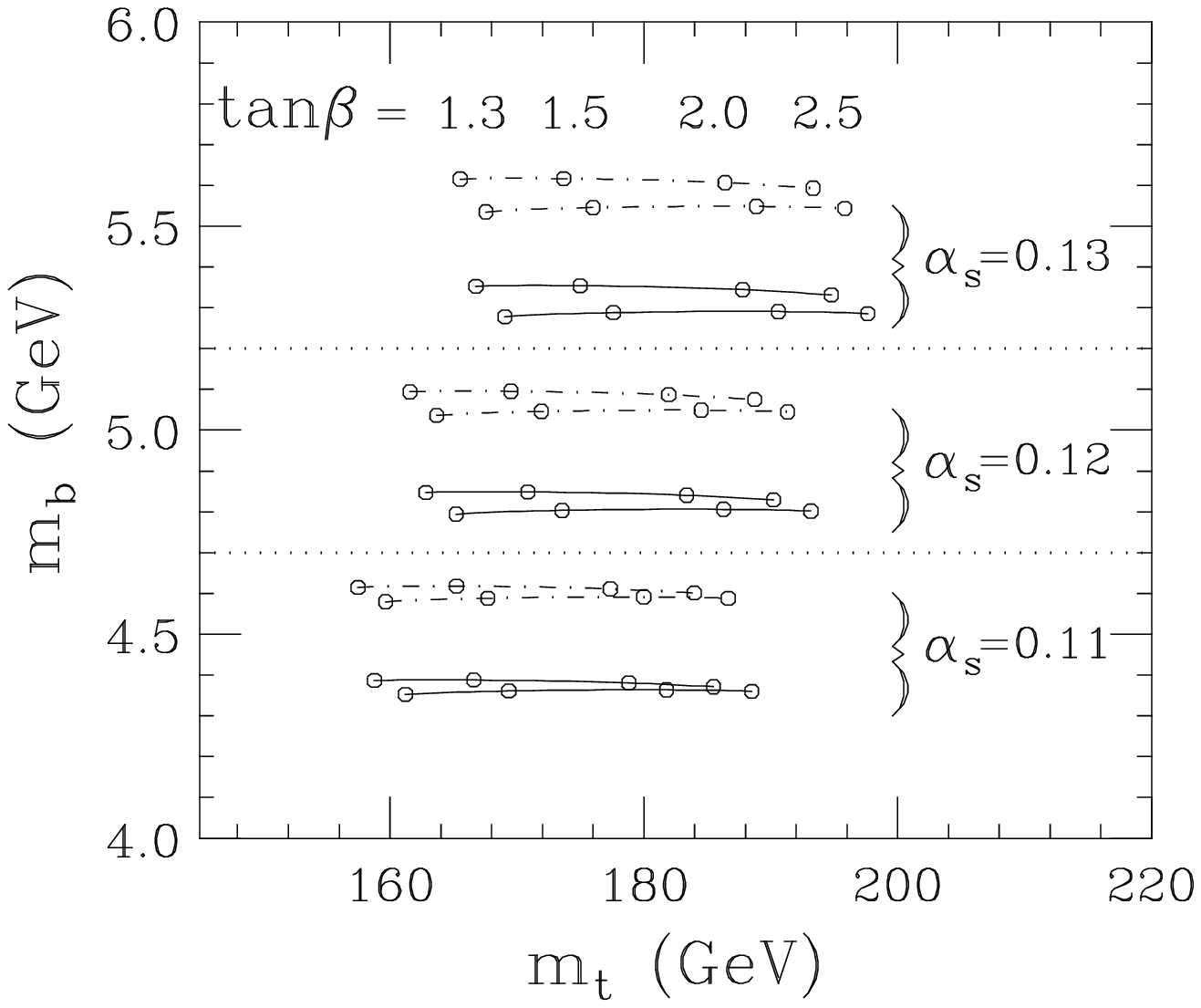}
\begin{center}
\parbox{5.5in}{
\caption[]{\small The bottom-quark mass vs.~the top-quark mass
for fixed values of $\alpha_s(M_Z)$ and various $\tan\beta$.
The solid lines correspond to $\hat\lambda_t(M_{\rm GUT})=3$
while the dashed lines correspond to $\hat\lambda_t(M_{\rm GUT})=2$.
The upper line in each pair corresponds to a light
supersymmetric spectrum with $M_0=M_{1/2}=80$ GeV. The lower line
in each pair corresponds to a heavy spectrum,
$M_0=1000$ GeV, $M_{1/2}=500$ GeV. The dotted lines delineate the
preferred region $4.7 < m_b < 5.2$ GeV. For $A_0=0$ and $\mu>0$.}}
\end{center}
\end{figure}

The study of gauge coupling unification has been carried out by many
groups beginning more than 20 years ago.  There has been a resurgence
during the last five years.  The analyses have become increasingly
more refined.  The most recent analyses use the full set of two-loop
RGE's to predict the strong coupling constant as a function of the
electroweak input parameters.  They pay proper attention to \ms-\ \dr
differences and treat the weak thresholds in different levels of detail.

Over time, the predicted value of the strong coupling constant has
increased markedly, in part due to the refinements mentioned above,
but more so from the fact
that the standard-model weak mixing angle, as determined by a
global fit to the data, has been steadily decreasing.  (This is
correlated with the increasing best-fit value for the top-quark mass.)

In this talk we take a closer look at supersymmetric unification.
We treat the supersymmetric threshold corrections in a complete
one-loop analysis.\footnote{See Ref.~\cite{Chankowski} for a similar
treatment of finite corrections to $\sin^2\theta_W$.}  Our work stands
in contrast to most previous studies, which are based on the ``leading
logarithm approximation."  This approximation involves taking the
standard-model value of $\sin^2\theta_W$ and adding the logarithmic
parts of the SUSY threshold corrections.  The approximation works well
if all of the SUSY particle masses are much greater than $M_Z$, in which
case the decoupling theorem implies that the finite effects of the
SUSY particles are negligible for all low-energy observables.

However, in realistic models it is not unusual for the supersymmetric
spectrum to contain light particles of order the $Z$-mass.  In this
case one cannot use the standard-model value of $\sin^2\theta_W$
as an input into a precision analysis.  This is because the quoted
value of $\sin^2\theta_W$ is the result of a fit to the data, assuming
that the standard model is correct.  The experimental analyses do not
include the finite SUSY corrections, which are different for each observable.
Therefore in our analysis, we use a single set of inputs in our calculation
of $\sin^2\theta_W$, namely, $\alpha_{EM}$, $G_F$,  $M_Z$, $m_t$, and the
parameters that describe the supersymmetric model.

We note that a careful evaluation of the weak mixing angle is
important for determining a precise prediction for $\alpha_s(M_Z)$.  Using
the one-loop RGE's and the condition of coupling unification, we find
that the three gauge couplings satisfy
\begin{equation}
{\beta_2-\beta_3\o \hat g_1^2(\mu)}\ +\ {\beta_3-\beta_1\o \hat g_2^2(\mu)}\ +
\ {\beta_1-\beta_2\o \hat g_3^2(\mu)} \ =\ 0\ ,
\end{equation}
where $\beta_i$ are the three beta functions, $d\hat g_i/dt = \beta_i
\hat g_i^3/16\pi^2$.  These relations imply that
\begin{equation}
{\delta\alpha_s\o\alpha_s(M_Z)} \ \simeq
\ \,{\delta\hat\alpha\o\hat\alpha} \ -
\ 7.5 \ {\delta\hat s^2\o\hat s^2} \ .
\label{delta_alpha}
\end{equation}
Hence, an error in the determination of $\hat s^2$ of 1\% leads to an
error in $\alpha_s(M_Z)$ of 7.5\%.

In this talk we use the full set of one-loop radiative
corrections to evaluate the \dr gauge and Yukawa couplings.  The \dr
couplings serve as the boundary conditions for the two-loop gauge and
Yukawa coupling renormalization group equations (RGE's), which
determine the couplings at very high scales.  In what follows we
use the full one-loop corrections at both the weak and GUT
scales to determine the regions of supersymmetric parameter space
that permit gauge and Yukawa coupling unification.

\pagebreak
\section{Calculation of $\hat s^2$}
\vspace{-.25cm}

Given the inputs $\alpha_{EM} = 1/137.036$,
$G_F=1.16639 \times 10^{-5}$ GeV$^{-2}$, and $M_Z=91.187$ GeV, as well
as $m_t$ and the parameters of the supersymmetric
model, we determine \cite{Degrassi} ($\hat c^2 = 1-\hat s^2$)
$$\hat\alpha = {\alpha_{EM}\over(1-\Delta\hat\alpha)}
\qquad\mbox{and}\qquad \hat c^2\hat s^2 = {\pi\,\alpha_{EM}\over
\sqrt2\,G_F\,M_Z^2\,(1-\Delta\hat r)}\ .$$
Here $\Delta\hat\alpha$ contains logarithms of the masses
of the charged particles, and
$$\Delta\hat r= \Delta\hat\alpha - {\hat\Pi_Z(M_Z^2)\over M_Z^2}
+ {\hat\Pi_W(0)\over M_W^2} + \mbox{vertex $+$ box}\ .$$
The vertex and box diagram contributions are the so-called
``non-universal" or ``non-oblique" corrections, and the remaining
corrections involve the real and transverse parts of the \dr gauge
boson self-energies.  The vertex
and box corrections vanish in the leading logarithm
approximation; the correction $\Delta\hat\alpha$ contains only
logarithms; and the $W$ and $Z$ self-energies contain
both logarithmic and finite corrections.

In our calculation of $\Delta\hat r$ we include the dominant two-loop
corrections (given in Ref.~ \cite{FKS}), which leads to a very precise
determination of $\hat s^2$.  Following Ref.~ \cite{Chankowski}, we
estimate the theoretical uncertainty in $\hat s^2$ to be about 1 part
in 10$^4$, while the experimental uncertainty (due to the uncertainty
in the determining the electromagnetic coupling at the $Z$-scale)
is 2.6 parts in $10^4$. Having determined $\hat\alpha$ and $\hat s^2$
precisely, we are in position to fix the boundary conditions for the
two-loop RGE's \cite {RGEs},
$$ \hat\alpha\equiv {\hat e^2\over4\pi},
\quad\hat s^2\qquad\longrightarrow\qquad
\hat g_1 = \sqrt{5\over3}{\hat e\over\hat c},\quad
\hat g_2 = {\hat e\over\hat s}\ ,$$
and accurately investigate gauge coupling unification.

In the following we assume that the SUSY masses unify at the
scale $M_{\rm GUT}$ (which is defined as the scale where $\hat g_1$
and $\hat g_2$ meet).  Therefore the supersymmetric model is parametrized
by a universal gaugino mass $M_{1/2}$, a universal scalar mass $M_0$, and
a universal trilinear scalar coupling $A_0$. The ratio of \dr vacuum
expectation values evaluated at the scale $M_Z$ is denoted $\tan\beta$.
We require the parameters to be such that electroweak symmetry is
spontaneously broken.  These conditions determine the value of $\mu^2$,
the supersymmetric Higgs  mass parameter, and $m_A$, the CP-odd neutral
Higgs boson mass, once we specify the sign of $\mu$.

\section{Prediction for $\alpha_s(M_Z)$}
\vspace{-.25cm}

As a reference point, we show in Fig.~2 contours of
$\alpha_s(M_Z)$ in the $M_0,\ M_{1/2}$ plane, with no GUT
thresholds, $\tan\beta=2$, $m_t=170$ GeV, and $A_0$=0.
We find $\alpha_s(M_Z)$ is large compared to the PDG value \cite{pdg}
$\alpha_s(M_Z)=0.117\pm0.005$, especially near $M_0,\ M_{1/2}
=100$ GeV.

The experimental uncertainty in the determination of $\alpha_s(M_Z)$ is
primarily due to the uncertainty in determining the electromagnetic
coupling at the $Z$-scale.  We use the value recently determined by
Eidelman and Jegerlehner \cite{Eidelman}.
Martin and Zeppenfeld \cite{Martin} and Swartz \cite{Swartz}
have also performed analyses to determine $\alpha(M_Z)$.
We show in Fig.~3 the differences in the determination of
$\alpha_s(M_Z)$ using the various values of $\alpha(M_Z)$.

As stated in the introduction, the finite corrections
can be significant when some of the superpartners have masses
of order $M_Z$.  This is illustrated
in Fig.~4, where we compare the value of $\alpha_s(M_Z)$ in the
leading logarithm approximation (LLA) with the value obtained in the
full calculation. In Fig.~4(a) the full and LLA curves converge
for large $M_{1/2}$ because the SUSY particles decouple.
In Fig.~4(b) the full and LLA curves
do not converge as $M_0$ becomes large.  This is because $M_{1/2}
= 60$ GeV, so the gauginos remain light for
arbitrarily large $M_0$.

To summarize our results for $\alpha_s(M_Z)$ in the absence of
GUT threshold corrections, we find $\alpha_s(M_Z) > 0.126$ for
squark masses less than 1 TeV, with $m_t=170$ GeV. For a SUSY
spectrum of 500 GeV or less, we have $\alpha_s(M_Z) > 0.130$.

If we require smaller values of $\alpha_s(M_Z)$  and a light
supersymmetric spectrum,
a GUT threshold correction is clearly needed.  We can parametrize
the GUT threshold correction by $\varepsilon_g$, where
$$ \hat g_3(M_{\rm GUT}) = \hat g_{\rm GUT}(M_{\rm GUT})\left(
1+\varepsilon_g\right)\ ,$$
and $\hat g_{\rm GUT}\equiv \hat g_1(M_{\rm GUT})=\hat g_2(M_{\rm GUT})$.
A smaller value of $\alpha_s(M_Z)$ requires $\varepsilon_g<0$.  In
what follows we examine the value of $\varepsilon_g$ in two SU(5)
GUT models.

In the minimal SU(5) model \cite{minSUfive}, the gauge coupling
threshold correction $\varepsilon_g'$ is given by
 \cite{minSUfive epsg}
\begin{equation}
\varepsilon'_g\ =\ {3g_{\rm GUT}^2\o40\pi^2}\,\log\left(M_{H_3}\o
M_{\rm GUT}\right)\ , \label{epsprime}
\end{equation}
where $M_{H_3}$ is the mass of the color-triplet Higgs particle that mediates
nucleon decay. From this expression, we see that $\varepsilon'_g < 0$ whenever
$M_{H_3} < M_{\rm GUT}$.  However, $M_{H_3}$ is bounded from below by proton
decay experiments. The $M_{H_3}$ mass limit is of the form \cite{Hisano}
$$ M_{H_3} > \M\ {|1+y^{tK}|\over\sin2\beta} f(\tilde w, \tilde d,
\tilde u,\tilde e) $$
where $\M$ is a nuclear matrix element, $y^{tK}$ parametrizes
the amount of third generation mixing, and $f$ is a function
of the wino, squark and slepton masses.

In Fig.~5  we show the minimum value for $M_{H_3}$ from the nucleon
decay constraint, for the conservative choices $\M=0.003$ GeV$^3$
and $|1+y^{tK}|=0.4$. We see that $M_{H_3}^{\rm min}>M_{\rm GUT}$
unless $M_0>500$ GeV and $M_{1/2} \ll M_0$. Thus, in most of the
parameter space, $\varepsilon_g'>0$.

In minimal SU(5), $\alpha_s(M_Z)$ is typically even larger than
it was without any GUT thresholds, as illustrated in Fig.~6.
The only exception occurs in the region $M_0 \gg M_{1/2}$, where the proton
decay amplitude is suppressed. In this region, with 1 TeV squark masses
and $m_t=170$ GeV, we find $\alpha_s(M_Z)$ as small as 0.123.
In fact, as long as $m_{\tilde q}\le1$ TeV,
$\alpha_s(M_Z)<0.126$ can only be obtained in the region
$M_0\simeq1$ TeV. For example, if $M_0\le500$ GeV,
$\alpha_s(M_Z)\ge0.127$.

The missing-doublet model is an alternative SU(5) theory in which
the heavy color-triplet Higgs particles are split naturally
from the light Higgs doublets \cite{missing-doublet}.  In this
model the GUT gauge threshold correction is given by \cite{Yamada}
\begin{equation}
\varepsilon_g^{\prime\prime}\ =
\ {3g_{\rm GUT}^2\o40\pi^2}\,\Biggl\{\log\left(M_{H_3}^{\rm eff}
\o M_{\rm GUT}\right) - {25\o2}\log5 + 15\log2\Biggr\}\ \simeq
\ \varepsilon'_g - 4\%\ .
\label{mdmodel}
\end{equation}
Thus, for fixed $M_{H_3}$, the missing-doublet model has the same
threshold correction as the minimal SU(5) model, minus 4\%.  In
eq.~(\ref{mdmodel}), $M_{H_3}^{\rm eff}$ is the effective mass that enters
into the proton decay amplitude, so the bounds on $M_{H_3}$ in the minimal
SU(5) model also apply to $M_{H_3}^{\rm eff}$ in the missing-doublet
model.

The large negative correction in eq.~(\ref{mdmodel}) is due to the mass
splitting in the {\bf 75} representation, and gives rise to much smaller
values for $\alpha_s(M_Z)$.  This is illustrated in
Fig.~7, where we show contours of $\alpha_s(M_Z)$ in the $M_0,\ M_{1/2}$
plane, with $M_{H_3}^{\rm eff}=M_{H_3}^{\rm min}$. The values
of $\alpha_s(M_Z)$ are somewhat low, but one can easily obtain larger
values, for example, by increasing $M_{H_3}^{\rm eff}$.

In Fig.~8 we illustrate the full range of $\alpha_s(M_Z)$ values in the
minimal SU(5) and missing-doublet models.  We show the allowed range of
$\varepsilon_g$ in the two models (setting $M_{H_3}^{\rm max}=10^{19}$ GeV),
together with the range of $\varepsilon_g$ which yields $\alpha_s(M_Z)=0.117
\pm0.01$. The missing
doublet model is mostly contained within the preferred region of
$\alpha_s(M_Z)$, while minimal SU(5) is almost entirely outside.

\pagebreak
\section{Yukawa unification}
\vspace{-.25cm}

In the final part of this talk, we investigate the possibility of
bottom-tau Yukawa coupling unification.  Our procedure is as follows.
We start with the experimental value for the $\tau$ pole mass $m_\tau=1.777$
GeV \cite{pdg}.  We convert it to the \dr running mass and evolve it up
to the $Z$-scale, where we apply the SUSY threshold corrections.  We
compute the \dr vev from the $Z$-boson mass,
$\hat v^2 = 4(M_Z^2+\hat\Pi_Z(M_Z^2))/(\hat g'^2+\hat g^2)$,
and use it to determine the \dr tau Yukawa coupling
$$\widehat\lambda_\tau = {\sqrt2\hat m_\tau\over \hat v\cos\beta}\ .$$
We then solve the RGE's to find $\widehat\lambda_\tau(M_{\rm GUT})$, and
set the bottom Yukawa coupling to
$$\hat\lambda_b(M_{\rm GUT}) = \widehat\lambda_\tau(M_{\rm
GUT})\left(1+\varepsilon_b\right)\ ,$$
where $\varepsilon_b$ parametrizes the GUT threshold correction.  Once
we have $\hat\lambda_b(M_{\rm GUT})$ we run everything back to the weak
scale and self-consistently determine the pole mass for the bottom quark.

Let us first examine the prediction for the bottom-quark pole mass with
$\varepsilon_b=0$.  Generally, the large value of the strong coupling
increases the bottom mass so much that the
prediction typically falls outside the
region determined by experiment \cite{pdg}, which we take to be
4.7 GeV $<m_b<5.2$ GeV.  Because Yukawa couplings enter the Yukawa
RGE's with the opposite sign from gauge couplings, large Yukawa
couplings help reduce the large $b$ mass. In particular, in the very
small $\tan\beta$ region, the top Yukawa coupling becomes large.  In
this infrared fixed point region the bottom mass can be less than
5.2 GeV.

We show in Fig.~9 the prediction for $m_b$ and $\alpha_s(M_Z)$, obtained
by setting the top Yukawa coupling to be
$\widehat\lambda_t(M_{\rm GUT})=3$.  We see that even with this
large top Yukawa coupling, which is on the verge of being non-perturbative,
$m_b$ is larger than 5.2 GeV unless $M_0$ or $M_{1/2}$ is greater than
about 1 TeV. One needs a GUT threshold correction to reduce $m_b$ with
a SUSY mass scale below 1 TeV.

The similarity between the curves for $\alpha_s(M_Z)$ and
$m_b$ in Fig.~9 illustrates the strong correlation between $\alpha_s(M_Z)$
and $m_b$. In fact, $m_b$ is far more sensitive to the gauge-coupling GUT
threshold correction than that of the bottom-quark Yukawa coupling.
Numerically, we find
$${\delta m_b\over m_b}\ \simeq\ 0.8\,\varepsilon_b\ +\ 8\,\varepsilon_g\ .$$
Hence, if we consider a GUT model where $\varepsilon_g$ is sufficiently
negative, the central value of $m_b$ can be obtained with
$\varepsilon_b=0$. This is shown in Fig.~10, where, for fixed
$\varepsilon_b=0$, we show the predicted value of $m_b$, assuming various
values of $\varepsilon_g$ that yield particular values of $\alpha_s(M_Z)$.
We show the results for $\widehat\lambda_t(M_{\rm GUT})=2$ and 3,
and for a small and large supersymmetric mass scale.
The figure shows that as long as $\varepsilon_g$ is such that
$\alpha_s(M_Z)\simeq0.12$, an acceptable value of $m_b$ is predicted for
$\varepsilon_b=0$, independent of the top-quark mass and the
supersymmetric mass scale.

\section{Conclusion}
\vspace{-.25cm}

In this talk we have presented results from a complete
calculation of the one-loop corrections to the masses, gauge, and
Yukawa couplings in the MSSM.  We have seen that such a calculation
allows us to reliably investigate various unified models to
see whether they are compatible with current experimental data.

In particular, we
found that the finite SUSY corrections, which are neglected
in the leading logarithm approximation, can substantially increase the
prediction for $\alpha_s(M_Z)$ when some of the SUSY partner
masses are lighter than or of order $M_Z$. In the minimal SU(5) model, we
found that $\alpha_s(M_Z)\,\roughly{>}\,0.14$ in the small $M_{SUSY}$ region,
$M_0\simeq M_{1/2}\simeq100$ GeV.  We also found $\alpha_s(M_Z)>0.123$ in
the region where the squark masses are below 1 TeV. In contrast, we showed
that the missing-doublet SU(5) model can accommodate much smaller values of
$\alpha_s(M_Z)$, such as $\alpha_s(M_Z)\simeq0.113$ for $M_0\simeq
M_{1/2}\simeq100$ GeV.

This work was supported
by the U.S. National Science Foundation under grant NSF-PHY-9404057.

\end{document}